\documentclass[sigconf, nonacm]{acmart}

\setcopyright{none}

\AtBeginDocument{%
  }

\usepackage{tikz}

\usepackage{placeins}
\usepackage{xcolor}
\usepackage{soul}

\soulregister\cite7
\soulregister\ref7
\soulregister\S0
\soulregister\emph1

\usepackage{adjustbox}
\usepackage{booktabs}
\usepackage{multirow}
\usepackage{flafter}
\usepackage{tabularx}
\usepackage{array}
\usepackage{longtable}
\usepackage{subcaption}

\usepackage{xfp}
\newcommand{\totalN}{22}

\newcommand{\pct}[1]{\fpeval{round(100*#1/\totalN, 0)}\%}

\usepackage{xurl}
\usepackage[most]{tcolorbox}

\usepackage{listings}

\usepackage{csquotes}

\newcounter{takeawaycnt} 
\definecolor{utkorange}{HTML}{FF8200}

\newtcolorbox{takeawaybox}{
    enhanced,
    breakable,
    rounded corners,
    colback=gray!10,        
    colframe=gray!10,       
    borderline west={4pt}{0pt}{utkorange}, 
    boxrule=0pt,
    sharp corners=northeast, 
    sharp corners=southeast,
    boxsep=2pt,             
    left=5pt,               
    right=3pt,              
    top=2pt,                
    bottom=2pt
}

\newenvironment{mytakeaway}
    {%
    \begin{takeawaybox}%
    \textbf{Lesson \thetakeawaycnt: }%
    \stepcounter{takeawaycnt}%
    }
    {%
    \end{takeawaybox}
    }

\usepackage{framed}
\usepackage{dirtytalk}
\definecolor{quote}{rgb}{0.98, 0.98, 0.98}
\definecolor{bar}{rgb}{0.0, 0.0, 0.0}

\newcommand{\longsay}[2]{%
  \def\FrameCommand{%
    \hspace{2pt}%
    {\color{bar}\vrule width 2pt}%
    {\color{quote}\vrule width 4pt}%
    \colorbox{quote}%
  }%
  \MakeFramed{\advance\hsize-\width\FrameRestore}%
  \begin{list}{}{%
    \setlength{\topsep}{0pt}
    \setlength{\leftmargin}{0pt}
    \setlength{\rightmargin}{0pt}
  }
  \item[]
    \say{\textit{#1}} (#2)%
  \end{list}%
  \endMakeFramed%
}

\let\savedsay\say
\renewcommand{\say}[1]{\savedsay{\textit{#1}}}

\RequirePackage{microtype}
\microtypecontext{spacing=nonfrench}

\RequirePackage{textcomp}
\RequirePackage{upquote}
\RequirePackage[all]{nowidow}

\emergencystretch=\maxdimen
\AtBeginDocument{%
  \hypersetup{
    colorlinks=true,
    citecolor=blue,
    linkcolor=blue,
    urlcolor=blue
  }%
}

\begin{document}

\title[A Multi-Month Study of Git Commit Signing]{A Multi-Month Study of Git Commit Signing}

\author{Abubakar Sadiq Shittu}
\orcid{0000-0002-4699-3015}
\affiliation{%
  \institution{University of Tennessee}
  \city{Knoxville}
  \state{TN}
  \country{USA}
}
\email{ashittu@vols.utk.edu}

\author{John Sadik}
\orcid{0009-0007-8988-5501}
\affiliation{%
  \institution{University of Tennessee}
  \city{Knoxville}
  \state{TN}
  \country{USA}
}
\email{jsadik@vols.utk.edu}

\author{Scott Ruoti}
\orcid{0000-0002-6917-4186}
\affiliation{%
  \institution{University of Tennessee}
  \city{Knoxville}
  \state{TN}
  \country{USA}
}
\email{ruoti@utk.edu}

\settopmatter{authorsperrow=3}

\renewcommand{\shortauthors}{Shittu et al.}

\begin{abstract}
Git commit signing, introduced in 2012, is one mechanism for establishing commit provenance in software supply chains, yet developer-controlled adoption remains rare and developers' experiences using it are understudied. To examine this experience, we conducted a three-month study with senior undergraduate and graduate computer science students ($n = 22$), whom we treat as proxies for junior developers. Participants configured commit signing independently, used it across four coursework projects, extended it to a second device, examined an external repository containing anomalous commits, and answered security-reasoning prompts. We found that while almost all participants successfully signed every commit and rated routine signing positively, many faced friction during setup, multi-device configuration, and repository verification. Despite signing all semester, they struggled to spot anomalous commits during verification, with over a quarter finding none. Additionally, nearly half expressed at least one misconception regarding signing guarantees or key management. Without isolating whether these difficulties stemmed from tooling, education, or understanding, we conclude that making signing easier is not enough to ensure effective security use. Rather, secure adoption also requires tools and education that support signature verification against authorized identities, interpretation of missing signatures and unknown keys, and correct reasoning about key-lifecycle operations.
\end{abstract}


\maketitle

\section{Introduction}
\label{sec:intro} 

Trustworthy software relies on the integrity of its source code, protection against unauthorized tampering, a secure build pipeline, and a dependable distribution mechanism~\cite{ishgair2026sok, souppaya2022secure,torresarias2019intoto,slsa2025threats}. This paper focuses on preventing and detecting unauthenticated modifications to source code.

Cryptographic commit signing is one possible defense against such changes~\cite{ishgair2026sok}. If a project only accepts commits signed with approved keys, it can detect cases where an attacker takes over a developer account or forges Git identity information without access to the developer’s signing key~\cite{holtgrave2025attributing,zhang2025pushing}. However, a valid signature is not enough on its own. An attacker may still use an unauthorized key while pretending to be another developer in the commit metadata~\cite{holtgrave2025attributing}. Also, a valid signature from an approved key does not guarantee that the signed code is benign. Commit signing therefore supports decisions about provenance and authorization, not about code correctness~\cite{ishgair2026sok}.

To be useful, verification must distinguish between unsigned commits, invalid signatures, valid signatures that cannot be linked to the claimed developer, and valid signatures from known but unauthorized keys~\cite{ishgair2026sok}. This only works well when developers sign consistently, protect and manage their keys, verify signatures against appropriate trust and authorization policies, and understand what the resulting verification states mean.

Despite these potential benefits, prior work shows that developers rarely sign their commits~\cite{holtgrave2025attributing,10.1145/3756681.3756959,zhang2025pushing,shittu2026analysis,mockus2026claimed}, and developers who adopt signing often use it inconsistently across projects~\cite{shittu2026analysis}. However, these measurement studies cannot reveal what developers experience while adopting it or whether sustained signing is accompanied by the verification ability and security understanding needed to use signatures effectively. In this paper, we examine this experience using a group of student developers and we do not attempt to identify a single cause of low adoption.

To this end, we conducted a semester-long study with senior undergraduate and graduate computer science students ($n = 22$), serving as a scoped proxy for junior developers entering the workforce. Participants independently configured commit signing, applied it to project submissions throughout the semester, and extended their setups to a second device. They also attempted to verify an external repository containing injected anomalies and answered prompts regarding commit-signing guarantees and key management. Because signing was required coursework, the study does not measure voluntary adoption or underlying motivations. Moreover, the study lasted only a few months, so long-term key management events, such as key revocation and expiration, remained out of scope. In summary, our work makes the following contributions:

\begin{enumerate}
    \item \textbf{Documented sustained signing and positive perceived usability.}
    Participants moved from successfully configuring commit signing to using it routinely throughout the semester. Most signed every observed commit ($n=18$, 81.8\%), and by the end of the semester, the cohort rated their experience with commit signing positively (SUS: $M = 77.61$).
    
    \item \textbf{Characterized the gap between sustained signing and effective security use.}
    Participants struggled to translate sustained signing into effective verification. On average, they identified only 4.77 of the 13 problematic commits ($Mdn = 3.50$). More than a quarter identified none ($n=6$, 27.3\%), while no participant identified every problematic commits. In addition, we found a similar gap in security understanding, with nearly half ($n=10$, 45\%) providing at least one misconception or incorrect answer about signing guarantees or key-lifecycle operations. 
\end{enumerate}

\noindent\underline{Significance:} These findings show that simply using commit signing does not guarantee security. While participants successfully signed commits throughout the semester, they persistently struggled to verify signatures and misunderstood fundamental security concepts. Our study does not isolate whether these specific challenges stem from confusing tools, poor training, or workflow. Regardless of the precise cause, usability improvements alone are unlikely to be sufficient. Secure adoption of commit signing requires the ecosystem to move beyond simplified key setup and provide tooling and training that actively support developers in interpreting commit verification states, navigating Git workflow exceptions, and understanding the precise boundaries of cryptographic guarantees.

\section{Related Work}
\label{sec:relatedwork}

\subsubsection*{Commit Signing Measurement}
\label{subsec:measurement}

Repository studies find limited developer-controlled signing, although estimates vary with sampling and the treatment of platform-generated signatures. Sharma et al.~\cite{10.1145/3756681.3756959} analyzed 60 repositories from four domains over five years and found that approximately 10\% of commits were verified, with variation across domains and Git clients. Holtgrave et al.~\cite{holtgrave2025attributing} analyzed 50,328 critical open-source projects and found that 95.4\% of users had never signed a commit and 72.1\% of projects had no signed commits. Only 2.0\% of users and 0.2\% of projects signed every commit.

Developer and ecosystem studies reveal further distinctions. Shittu et al.~\cite{shittu2026analysis} analyzed 71,694 developers across approximately 16 million commits and 874,198 repositories. Excluding platform-generated signatures, fewer than 6\% had signed locally. Among those who did, signing rarely persisted across time and repositories, approximately one in eight signatures failed verification because the public key was not uploaded, and more than one-quarter of users with registered keys had an expired or otherwise unusable key. Mockus~\cite{mockus2026claimed} found signatures in 17.59\% of 5.87 billion commits and separated individual keys from organizational and continuous integration keys. Thus, repository prevalence, personal adoption, sustained use, and successful verification are distinct, and aggregate rates may reflect automation rather than individual practice. These studies measure observable adoption but not developers performing signing tasks. Zhang et al.~\cite{zhang2025pushing} provide the closest account of developer experience through interviews with 17 open-source contributors
and an analysis of 12.5 million commits. Participants had limited awareness of impersonation, underestimated its severity, and found key management across machines burdensome. This study captures reported perceptions and barriers but
does not observe the same developers through setup, sustained use, second-device use, and repository verification.

\subsubsection*{Software Provenance, Trust, and Identity}
\label{subsec:provenance}

A valid commit signature shows that a commit has not changed and was signed with the corresponding private key. It does not prove the signer's identity, authorization, or that the code is safe~\cite{holtgrave2025attributing,zhang2025pushing}.
Because Git identities are freely configurable, attackers can easily impersonate legitimate developers, spoof contributions, or hijack project reputation in the vast majority of open-source workflows~\cite{holtgrave2025attributing,zhang2025pushing}. Repository trust therefore requires identity binding, contributor policies,
and signature verification.

Trust also requires complete history and verifiable platform actions. Torres Arias et al.~\cite{torresarias2016omitting} demonstrated attacks that omit security patches, reintroduce vulnerable code, or present inconsistent histories, and proposed signed records for detection. Afzali et al.~\cite{afzali2018legitimate} made actions performed through web-based Git interfaces verifiable while remaining compatible with Git signatures.

Commit signing is only one part of software provenance. In-toto uses signed metadata to verify source, build, test, packaging, and release steps~\cite{torresarias2019intoto}. Sigstore uses short-lived certificates
linked to existing identities and transparency logs~\cite{newman2022sigstore}, while Speranza adds signer privacy~\cite{merrill2023speranza}. The latter two reduce dependence on long-lived developer-managed keys. Organizational context
adds further requirements. Interviews with 18 practitioners from 13 organizations identified technical, organizational, and human challenges, differing views about software signing's importance, and dependencies on infrastructure, policy, and automation~\cite{kalu2025industry}. Commit signing is therefore one provenance signal whose value depends on identity,
history integrity, verification, and organizational controls. Individual Git users do not fully represent mature supply chains that also use continuous integration, centralized key custody, access controls, compliance requirements, and formal review.

\subsubsection*{Usable Cryptographic Key Management}
\label{subsec:keymanagement}

Key management includes generation, storage, identity verification, use across devices, recovery, rotation, revocation, and retirement, requiring a balance among availability, recovery, and protection from unauthorized access~\cite{blessing2025recovery}. Users have long struggled with these responsibilities. Most participants could not use PGP securely despite its graphical interface~\cite{whitten1999johnny}. Automated certificate creation and distribution reduced routine effort but left users responsible for interpreting key changes~\cite{garfinkel2005johnny2}. Users preferred encryption integrated into familiar applications~\cite{atwater2015leading}, and comparisons showed that usability also depends on the underlying key management scheme~\cite{ruoti2018comparative}. Technical experience did not eliminate these problems: 87\% of developers and administrators believed they had issued an OpenSSL certificate, but only 45\% had done so~\cite{ukrop2018johnny}, and knowledgeable users struggled to deploy HTTPS~\cite{krombholz2017https}. Conversely, Let's Encrypt and Certbot show that automation and appropriate defaults can improve certificate management~\cite{tiefenau2019certbot}.

Key possession must also be connected to identity. Conventional fingerprints are vulnerable to comparison errors~\cite{dechand2016fingerprints}, with attack success ranging from 6\% to 72\% across designs~\cite{tan2017unicorns}. Instruction improved completion of authentication ceremonies but did not ensure understanding of their security guarantees~\cite{vaziripour2017alice}. CONIKS makes identity and key associations auditable, although systems must still distinguish legitimate key changes from attacks~\cite{melara2015coniks}.

Using one private key across devices simplifies identity management but increases exposure, while separate keys require additional distribution and removal. Cloud key stores support multiple devices but introduce trust in the service~\cite{kurnikov2018keys}. Loss requires recovery or replacement, whereas theft also requires revocation and notification. Users report lost or nonfunctioning recovery codes and misunderstand what providers can restore~\cite{holtervennhoff2024recovery}, reflecting tradeoffs among recovery, provider trust, and account takeover~\cite{blessing2025recovery}. Protected cloud infrastructure can support secure recovery but is rarely available to individual developers~\cite{connell2024recovery}.

Software signing faces the same lifecycle challenges. While The Update Framework supports recovery from compromise through separated roles, threshold signatures, and replaceable trust metadata~\cite{samuel2010survivable}, and dedicated policies and tools can improve software signing quantity and quality~\cite{schorlemmer2024signing}, enterprise practitioners still struggle with organizational key custody, integration, and recovery~\cite{kalu2025industry}. Moreover, newer systems retain integration and component maturity problems~\cite{kalu2026johnny}. Although Git commit signing requires developers to handle these same key management responsibilities, the specific challenges they encounter in practice remain largely unexamined.

\subsubsection*{Student Proxies}
\label{subsec:student}

Falessi et al.~\cite{falessi2018empirical} argued that the suitability of student proxies depends on the target population, study context, and relevant experience. Salman et al.~\cite{salman2015students} found similar student and professional performance in a controlled study of an unfamiliar development approach. Naiakshina et al.~\cite{naiakshina2020students} found similar relative effects of security prompting and framework support, although company developers performed better overall. Taken together, these findings support task-specific use of student proxies but do not establish equivalence in professional settings.

\section{Gap and Research Questions}
\label{sec:rq}

\S\ref{sec:relatedwork} documents low commit-signing adoption, the importance of identity binding, and persistent usability problems in key management. However, to our knowledge, no prior study has followed the same Git users through initial setup, sustained use, multi-device configuration, signature verification, and security reasoning over several months. To address this gap, we follow prior literature in using advanced computer science students as scoped proxies for junior developers and ask:

\begin{description}
\item[RQ1] What do developers experience when adopting and sustaining Git commit signing over time?
\item[RQ2] To what extent is sustained use of Git commit signing accompanied by the verification ability and security understanding needed to use it effectively as a security mechanism?
\end{description}
\section{Methodology}
\label{sec:method}

\subsection{Participants}
\label{subsec:participants}

\subsubsection{Study setting.} We embedded an IRB approved, semester long study of Git commit signing into a required applied cryptography course for senior and graduate computer science students at a public university in the United States. This setting provided a technically sophisticated participant pool whose signing behavior could be studied during normal coursework rather than in a laboratory.

\subsubsection{Proxy scope.} We selected this population deliberately because they bring a relevant technical background, including version control fluency, cryptography coursework, and practical implementation skills. However, they are not yet professional developers with years of signing practice. We treat this population as a proxy for \emph{junior} developers rather than developers broadly. Participants are seniors and graduate students who will enter the workforce within the next year or two, and many were already working as junior developers at the time of writing.

Although recruiting practicing junior developers would have provided a closer match to our target population, we did not have the corporate connections or funding required to recruit and retain a sufficient sample for a multi-month study. Recruiting professional developers for a study of this length would also have been difficult, even with additional resources. Moreover, senior developers, and even junior developers at companies with substantial onboarding, training, or IT support, may face different barriers than those we observed in this study. We return to this distinction in \S\ref{subsec:limitation} and view replication with professional developer populations as necessary future work.

\subsubsection{Background.}Before entering the course, participants had completed an Introduction to Cybersecurity prerequisite. During the course, they received theoretical and practical grounding in the cryptographic primitives that underpin commit signing, including the SHA hash family, message authentication codes, RSA, Diffie Hellman key exchange, and public key infrastructure. They also implemented cryptographic primitives, including AES according to FIPS~197 and RSA from scratch, and studied PKI usability and secure authentication. Consequently, the signing difficulties observed in this study are unlikely to result solely from unfamiliarity with digital signatures.

\begin{table}
    \centering
    \scriptsize
    \caption{Consolidated study design.}
    \label{tab:study-design}
    \setlength{\tabcolsep}{2pt}
    \renewcommand{\arraystretch}{1.15}
    \begin{tabularx}{\linewidth}{
        @{}
        >{\raggedright\arraybackslash}p{0.16\linewidth}
        >{\raggedright\arraybackslash}p{0.18\linewidth}
        >{\raggedright\arraybackslash}p{0.20\linewidth}
        >{\raggedright\arraybackslash}X
        >{\centering\arraybackslash}p{0.11\linewidth}
        @{}
    }
        \toprule
        \textbf{Phase} &
        \textbf{Assignment} &
        \textbf{Task} &
        \textbf{Data and Instruments} &
        \textbf{Results} \\
        \midrule

        First Setup &
        Part~1 &
        Set up signing keys and create and verify a first signed commit
        (\S\ref{procedure_1}) &
        Repository and \texttt{.git} history, public key,
        completion time, SUS and ASQ, and written self-reports. &
        \S\ref{sec:results-first-setup} \\
        \midrule

        Sustained Use &
        4 interim projects; Part~2, Section~3 &
        Sign commits across four coursework projects
        (\S\ref{procedure_2}) &
        \texttt{.git} archives, public key, semester-end
        written self-reports, and Semester-end SUS. &
        \S\ref{sec:longitudinal} \\
        \midrule

        Second Device &
        Part~2, Section~1 &
        Set up signing and create and verify a signed commit on a second device
        (\S\ref{procedure_3}) &
        \texttt{.git} history, completion time, written
        self-reports, and ASQ. &
        \S\ref{sec:second-device} \\
        \midrule

        Commit Verification &
        Part~2, Section~2 &
        Inspect a repository and identify commits with injected anomalies
        (\S\ref{procedure_4}) &
        Written self-reports containing flagged commits,
        rationales, ASQ, completion time, and tools used. &
        \S\ref{sec:finding-bad-commit} \\
        \midrule

        Security Understanding &
        Part~2, Section~4 &
        Answer five open-ended prompts about signing security and key management
        (\S\ref{procedure_5}) &
        Five open-ended prompts and written self-reports. &
        \S\ref{sec:understanding} \\

        \bottomrule
    \end{tabularx}
\end{table}

\subsubsection{Recruitment and consent.}Of 70 students enrolled in the class, only $n=22$ (31\%) provided informed consent to have their data included in this research. All students completed the course project as part of normal instruction; research participation was opt-in.  We refer to the 22 participants as P1 through P22. Table~\ref{tab:demographics} in Appendix~\ref{sec:supplementary-results} summarizes their demographics, and Appendix~\ref{ethics} describes the study's ethical considerations.

\subsection{Procedures}
\label{sec:procedures}

The study comprised two graded course assignments and four interim coursework projects conducted over three months. Together, these activities formed the study phases summarized in Table~\ref{tab:study-design}. All qualitative data were collected as written self-reports from the course assignments. We describe the procedure below.

\subsubsection{Initial Setup}
\label{procedure_1}

The first phase asked participants to configure cryptographic keys and Git commit signing independently, without step-by-step guidance from course staff. Each participant initialized a local repository, created a plain-text file containing their university ID, committed and pushed the file in a digitally signed commit, and verified its signature. Participants could use any tools, remote services, development environments, or public resources, reflecting how developers may adopt an unfamiliar security practice in real-world settings.

We deliberately withheld institutional guidance to preserve ecological validity, reflecting how developers must navigate unfamiliar cryptographic tooling without a predefined playbook. Providing tutorials would have reduced the very friction points the study aimed to observe. At the same time, to balance realism with ethical considerations in a classroom context, we implemented a tiered support policy. Participants were encouraged to work independently for at least 90 minutes before requesting hints, and up to three hours before seeking a full walkthrough. Any assistance received had to be disclosed in the written self-report so that it could be accounted for in the analysis.

Participants submitted their Git repositories, including the \texttt{.git} directory, along with their public keys. These materials allowed us to cryptographically verify commit signatures and inspect commit histories. Participants also submitted written self-reports describing the total time spent, the steps followed, the tools used or abandoned, and the information sources consulted. These written self-reports captured not only the final outcome but also the process through which participants arrived at a working setup. As part of the written self-report, participants also completed the System Usability Scale (SUS)~\cite{brooke1996sus} and the After Scenario Questionnaire (ASQ)~\cite{lewis1991psychometric}. These provided quantitative measures of perceived usability and satisfaction, enabling systematic comparisons across participants and, in the case of the SUS, multi-month tracking over the course of the semester.

\subsubsection{Sustained Use Across the Semester}
\label{procedure_2}

Commit signing was not treated as a single assignment. After the initial setup, participants were asked to sign commits across four coursework projects spanning three months of the semester. For each submission, they uploaded to Canvas a zip archive containing the project files, the \texttt{.git} directory, and any public keys needed to verify the signatures. These artifacts allowed us to determine whether signing continued across projects and how consistently participants applied it.

The instructions were intentionally vague and open ended so that we could examine how participants interpreted the requirement, specifically whether they signed only the final commits associated with graded submissions or consistently signed all commits in each repository. We graded only the submitted project and did not penalize unsigned commits, in order to avoid influencing behavior and to observe natural signing practices under minimal enforcement. Notably, this window spans three months of sustained use; it does not extend to key-lifecycle events that typically unfold over longer periods, such as key revocation or expiration, which we note as a limitation of our study (see \S\ref{subsec:limitation}).

At the end of the semester, participants submitted written self-reports covering key aspects of their workflow, including how their signing key was stored and secured, whether their workflow changed over time, and whether they intended to continue using commit signing beyond the course. These written self-reports were graded for completion only. We also administered the SUS a second time, enabling a pre/post comparison of usability perceptions from initial setup to end-of-semester use.

\subsubsection{Second Device Setup}
\label{procedure_3}

At the end of the semester, after the final project milestone, participants completed a portability task designed to reflect a common real-world scenario: extending a signing identity to a second device. Participants were required to produce a signed commit on a different device type from the one used in the initial setup. If they used a personal computer, they were given access to a university computing workstation, and vice versa if they started with one. The commit included a plain text file indicating which machine was used in each part, and participants were required to verify that the resulting signature was valid. Participants who had lost access to their Part 1 repository were permitted to retrieve it from their prior course submission before proceeding, allowing us to compare the \texttt{.git} log between both setups for the same user.

The task was intentionally open-ended, with no prescribed implementation strategy, to observe how participants approached cross-device key management without guidance. Participants chose between transferring their existing private key to the second device or generating a new keypair and managing multiple signing identities. These approaches have different security implications, and a goal of this phase was to understand which strategy users naturally adopt and why.

Participants submitted written self-reports describing their chosen approach, rationale, steps taken, failures encountered, tools used or abandoned, and information sources consulted. They also completed a post-task ASQ specific to the cross-device signing experience, rating their satisfaction with the ease of setup, time required, and quality of available support documentation.

\subsubsection{Commit History Verification}
\label{procedure_4}

To evaluate participants' ability to detect malicious activity through signature verification, we designed a trust assessment task that placed participants in a realistic scenario, acting as developers evaluating whether to incorporate an open source library into a work project, knowing that some repositories may contain malicious code inserted by attackers. We constructed a repository of 20 commits and deliberately injected 13 anomalous commits spanning five categories, including commits from an unknown developer, timestamp manipulation, author and committer identity mismatches, use of an incorrect signing key, and missing signatures from known developers. Participants were provided the repository URL and identity material for two authorized developers, including their names, email addresses, and SSH key fingerprints, and were asked to determine whether the repository was safe to use.

This design allowed us to measure three things: how many and which
categories of anomalies participants could detect, what strategies and
tools they used to conduct the analysis, and where the process broke
down. Participants submitted written self-reports identifying which commits they flagged and why, how long the task took, what steps and tools they used, and which aspects were easiest or hardest. They also completed an ASQ specific to this task.

\subsubsection{Security Reasoning}
\label{procedure_5}

To assess participants' security understanding of Git commit signing beyond procedural knowledge, we collected written self-reports in response to five open-ended prompts at the end of the assignment. The prompts were explicitly framed as having no single right or wrong answer, and participants received full credit regardless of the content of their written self-reports, encouraging candid answers rather than performance. The five questions covered: (1) the security benefits and drawbacks of commit signing compared to unsigned commits, (2) the security implications of synchronizing an existing signing key to a new device versus generating a new key per device, (3) the workflow and security concerns associated with private key loss, (4) the workflow and security concerns associated with private key theft, and (5) how an attacker might compromise a widely used open source repository and what practices could prevent such an attack. Together, these questions probed three levels of understanding: \textit{conceptual} understanding, referring to what signing provides and the trade-offs of key management; \textit{operational} understanding, referring to key lifecycle management and the distinction between key loss and key theft; and \textit{threat modeling}, referring to participants' ability to reason about realistic attack scenarios and potential defenses.

\subsection{Data Analysis}
\label{sec:analysis}

\subsubsection{Quantitative Methods}
\label{sec:quan}

We computed task-completion rates, problematic commit detection rates, time-on-task, participant-level distributions, and responses to the SUS and ASQ. SUS scores were calculated using Brooke's standard scoring procedure~\cite{brooke1996sus} and interpreted using Bangor et al.'s~\cite{bangor2008empirical} acceptability ranges.

For continuous outcomes, we report the sample size, arithmetic mean, standard deviation, median, and interquartile range where informative. Because participant-reported completion times were positively skewed, we report geometric means together with medians and interquartile ranges, along with 95\% confidence intervals for means where available.

The study used a fixed sample of consenting students from an existing course, so an a priori power analysis was not used to determine the sample size~\cite{lakens2022sample}. Given the resulting sample size ($n = 22$), all inferential analyses are exploratory, and the reported two-sided $p$-values are interpreted cautiously. In particular, a non-significant result is described as no statistically significant difference or association observed in this sample, rather than as
evidence of equivalence or the absence of an effect~\cite{lakens2018equivalence}. 

We used Mann--Whitney $U$ tests~\cite{mann1947test} for independent group comparisons and Wilcoxon signed-rank tests~\cite{wilcoxon1945individual} for paired comparisons across time points, reporting effect size ~\cite{fritz2012effect}. We used Pearson's $r$ and Spearman's $\rho$ to examine relationships among completion time, usability, signing behavior, and detection performance. Complete descriptive statistics and full reporting for every inferential analysis---including the analyzed group sizes, relevant descriptive summaries, test statistics, $p$-values, effect sizes, and confidence intervals where available---are provided in Appendix~\ref{sec:supplementary-results}, Tables~\ref{tab:descriptive-results},
\ref{tab:inferential-adoption}, and \ref{tab:inferential-verification}.

\subsubsection{Qualitative Methods}
\label{sec:qual}

To analyze the open ended self-reports, two researchers collaboratively coded the responses using an open-coding approach~\cite{holton2007coding,strauss1990basics} to identify concepts directly from the data. Each code represented a unique segment or concept derived from one or more excerpts rather than a predetermined category. During this process, the researchers employed the constant comparative method~\cite{glaser1965constant}, continually comparing new excerpts against previously coded material to create, refine, merge, or split codes as needed. Both researchers were present during all stages of the coding process and discussed every coding decision as it arose. Because we resolved all disagreements through discussion until we reached full agreement, we did not calculate inter-rater reliability. 

For the security understanding self-reports, the same collaborative procedure was used to identify recurring reasoning patterns. In a separate evaluative step, we classified each answer as \emph{correct}, \emph{misconception}, or \emph{wrong} using our study specific operational definitions. A \emph{correct} answer was technically accurate, even if brief. A \emph{misconception} contained some technically correct reasoning but also an identifiable conceptual misunderstanding~\cite{national1997science}. A \emph{wrong} answer reached a fundamentally incorrect conclusion without demonstrating the relevant correct understanding. This classification assessed technical accuracy rather than completeness, as the primary goal was to identify misunderstandings that could carry security consequences rather than to measure the full depth of participants' knowledge. The details about full codebook is available in \S\ref{sec:open-science}.

\subsubsection{Data exclusions and robustness checks.} For the initial setup data, two participants were excluded from the SUS analysis: P18, due to missing responses across several items, and P19, due to the use of an incorrect Likert scale that rendered their submission incompatible with the rest of the dataset. However, their remaining data were retained, as their qualitative responses and observed task behaviors continued to provide useful insights. For the end-of-semester SUS analysis, P5 entered an out-of-range value for one item. We retained the response in the primary analysis after capping it 
at the maximum permitted scale value; that is, a response of 6 was treated as 5. As a robustness check, we recalculated the SUS results after excluding the adjusted response entirely. All substantive results remained unchanged.
\section{Findings: Initial Setup}
\label{sec:results-first-setup}

All participants ($n=22$) produced a signed commit without reaching out to the instructor,but the process was rarely immediate. We observed a clear trial-and-error pattern. Although the task required adding a single file, which ideally would be a single commit, if we consider commit count as a proxy for rework, \pct{8} of participants ($n=8$) recorded multiple attempts in their \texttt{.git} logs. The primary behavioral differentiator was the presence of technical hurdles.  Among participants with valid time reports, those who reported at least one pain point took significantly longer to complete the task than those who encountered none (geometric mean: 57.74 vs.\ 19.79 minutes; Appendix~\ref{sec:supplementary-results}, Table~\ref{tab:inferential-adoption}). This suggests that reported completion time was a sensitive behavioral marker of setup difficulty.

Participants first selected tools and information sources to complete the setup. Most chose GnuPG (GPG) ($n=16$, \pct{16}), while a smaller group used SSH keys ($n=6$, \pct{6}). No participant reported prior experience with commit signing, which helps explain both the dominance of GPG—reflecting its prominence in available documentation—and the reliance on external resources. All participants consulted at least one external source, indicating that commit signing is not easily self-discoverable. As P9 noted, \longsay{The hardest part is discovering how to sign a commit in the first place. The process isn't self-contained within Git; you need to go out of your way to find, install, and configure external tools like GnuPG before you can even begin.}{P9} GitHub’s official documentation ($n=16$) was the most commonly used and was generally rated positively for its clarity and completeness (Table~\ref{tab:info-sources}). We next examine the specific pain points and errors that emerged from this reliance on external guidance.

\subsection{Pain Points}

\subsubsection{Schema Discrepancies in SSH Verification}
\label{subsubsec:allowed-signers}

Participants using SSH signing ($n=4$: P2, P4, P5, and P7) encountered difficulties with the \texttt{allowedSignersFile}—a local configuration file required by Git to map developer identities to their trusted SSH public keys during verification. The primary issue was a subtle schema mismatch between standard SSH public key formats and Git’s expected structure for this file.

This mismatch was exacerbated by limited diagnostic feedback: when the format is incorrect, Git provides no specific error message, often resulting in an unsigned commit or a generic verification failure. As a result, participants frequently engaged in misdirected debugging, including attempts to fix unrelated issues such as line-ending inconsistencies, before identifying the structural requirement of the file. As P2 described: \longsay{I realized that \textup{\texttt{allowedSignersFile}} should have the format \textup{\texttt{<email> <key\_type> <key>}}, not the \textup{\texttt{<key\_type> <key> <email>}} format in the \textup{\texttt{id\_rsa.pub}} file. After changing the file, a final check verified that the signature was correct.}{P2}

P2 further characterized this as a design issue rather than a user error: \longsay{I find it unintuitive that while \textup{\texttt{ssh-keygen}} generates a file of a specific format, Git expects a file of a different format, although the contents are the same.}{P2}

\subsubsection{Email Mismatch}
\label{subsubsec:email-mismatch}

Participants ($n=2$, P11, P16) discovered mid-task that the email address used during GPG key generation must match the email registered with their GitHub account. Neither the key generation wizard nor the GitHub interface proactively communicates this constraint. 

\longsay{I also had an issue with the GitHub account not having the same email as the one I used for GPG, so I had to repeat the process \dots I would maybe include some warnings about the email and making sure it was the same as the git and or github email during the gpg installation.} {P11}

\subsubsection{GPG Passphrase Prompt Failure on macOS}
\label{subsubsec:pinentry}

Participants on macOS ($n=4$: P6, P8, P17, P18) encountered failures in which GPG could not prompt for the key passphrase within the terminal, causing signing operations to fail with either no visible prompt or a generic error. This issue is related to the requirement for the \texttt{pinentry-mac} helper, which renders a GUI passphrase dialog on macOS but is not included in the primary GitHub GPG setup instructions. P8 described the resulting error and resolution:
\longsay{\textup{\texttt{error: gpg failed to sign the data / fatal: failed to write commit object}}. To resolve this error, I configured GPG to use a PIN entry program that could display a GUI prompt for the passphrase.}{P8}

P8 further noted that the failure was difficult to diagnose:
\longsay{The error message didn't clearly indicate what was wrong. For me, it wasn't obvious that the issue was related to GPG prompts for passphrases.}{P8}

P6 similarly characterized the issue as a known but underdocumented platform bug, noting \enquote{a known bug with macOS GPG GUI} that required consulting a community forum to resolve.

\subsubsection{Git Not Configured to Use the Signing Key}
\label{subsubsec:git-config}

Participants ($n=6$: P5, P7, P8, P10, P13, P18) encountered issues where Git did not use their newly generated signing key, either because \texttt{git config user.signingkey} had not been set or because \texttt{git config gpg.format} was not configured for SSH users, causing Git to default to searching for a GPG key by default.

P10 made a related assumption that also proved incorrect---that Git would automatically associate the new GPG key because the same name and email had been used in both \texttt{gpg --full-generate-key} and the existing Git configuration: \longsay{I tried committing which of course didn't work [\ldots] I thought since I used the same name and email as my git config I wouldn't need to specify the key to git but that didn't work, so I used \textup{\texttt{git config ---global user.signingkey}} to add the ID.}{P10}


\subsubsection{Platform-Specific and Tooling Incompatibilities}
\label{subsubsec:platform}

A small subset of participants ($n=3$) encountered platform-specific or tooling-related constraints. P14 reported that GitHub Desktop does not support signed commits, requiring a switch to terminal-based workflows:
\longsay{The GitHub desktop app does not support working with keys \dots, so I was forced to switch to using the terminal and Windows Powershell for the remainder of the process.}{P14}

This constraint introduced a shift from GUI-based to command-line workflows during the task.

This finding also suggests a potential divergence in how participants might approach commit signing outside the study context. Without the requirement to submit signed commits locally, users of GUI-based clients may rely on GitHub’s web-based signing mechanisms rather than configuring local signing keys, which correspond to different security assumptions~\cite{shittu2026analysis}.

Across these cases, participants highlighted a lack of integration between Git and its signing dependencies as a recurring source of friction. P4 noted:
\longsay{I would make setting up commit signing a core feature of git, rather than using a mix of tools to get it working, like during git’s initial setup.}{P4}

\subsection{User Perceptions}

Despite the technical hurdles described above, subjective perceptions of the system remained generally positive (ASQ: $M = 6.12$; SUS: $M = 76.62$; Table~\ref{tab:descriptive-results}). Both measures were negatively correlated with perceived completion time (Appendix~\ref{sec:supplementary-results}, Table~\ref{tab:inferential-adoption}), indicating that longer experiences were associated with lower perceived usability, even when overall scores remained relatively high.

Participants also reported that the task increased their understanding of the underlying signing process. As P21 noted:
\longsay{I’m very familiar with Git and GitHub (my account is over seven years old), but I had never used GPG before, so this was a new experience. In the past, I relied on GitHub Desktop without understanding the underlying cryptography. This project helped bridge that gap by making the signing workflow and its security properties much clearer.}{P21}

\begin{mytakeaway}
All participants successfully set up Git commit signing, but many encountered configuration or environment specific failures. Participants who reported problems took longer to finish, and longer setup times were linked to lower usability ratings. Even so, participants generally rated the experience positively.
\end{mytakeaway}

\section{Findings: Semester-Long Usage}
\label{sec:longitudinal}

\begin{figure}
  \centering
  \includegraphics[width=\linewidth]{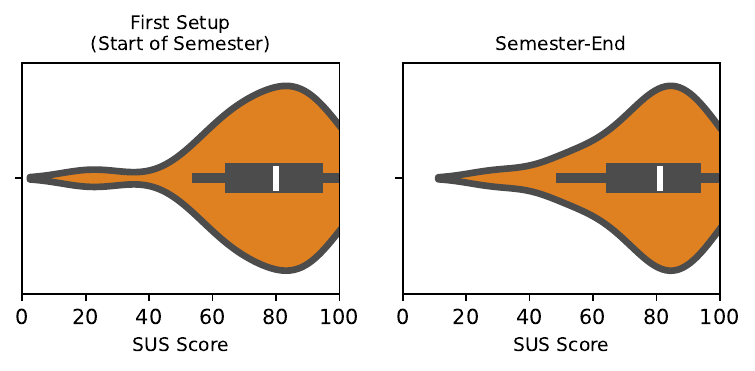}
  \caption{Distribution of SUS scores at initial setup and at semester-end after sustained use of commit signing.} 
  \Description{Distribution of SUS scores at first setup and semester-end.}
  \label{fig:sus-violin}
\end{figure}

As shown in Figure~\ref{fig:sus-violin}, participants reported a positive experience at semester-end (mean = 77.61, 95\% CI [69.9, 85.3], SD = 17.4, $n = 22$), corresponding to an adjective rating of \textit{Good}. Among participants with scores at both time points, we observed no statistically significant change in SUS scores between initial setup and semester-end (Appendix~\ref{sec:supplementary-results}, Table~\ref{tab:inferential-adoption}). This stable quantitative assessment was accompanied by several recurring themes.

\subsubsection*{Commit Signing Was Not Intuitively Understood as a Per-Commit Practice}

We found that most participants ($n = 18$) signed every observed commit. Interestingly, four participants (P3, P5, P14, and P16) did not do so consistently (57--87\%). This suggests that some participants did not naturally interpret signing as a per-commit practice. This may reflect limitations in instructional framing, as instructions did not explicitly require all commits to be signed. More broadly, this points to a potential conceptual gap between casual and professional signing norms. Notably, there was no statistically significant difference in signing consistency between participants with higher and lower SUS scores (Appendix~\ref{sec:supplementary-results}, Table~\ref{tab:inferential-adoption}). The small number of participants who did not sign every observed commit limits the strength of this comparison.

\subsubsection*{Effortless Daily Use and Workflow Stability}

Despite rating the initial setup as usable, participants drew a sharp distinction between setup and ongoing use. All but one participant ($n=21$) described the setup as considerably harder than daily signing, while nearly all ($n=20$) characterized routine signing as seamless or automatic once configured. This pattern held regardless of operating system, key type, or SUS score. 

\longsay{Setting it up was way way harder. Because once it was set up, all I had to do was commit as normal, but enter a passphrase every so often.}{P16}

\longsay{The initial setup took around 1.5--2 hours due to troubleshooting, new concepts, and new tools\ldots\ After setting it up, signing commits was easy to do.}{P8}

We also found that participants reported making no changes to their signing
process throughout the semester ($n = 16$), indicating rapid habit
formation after initial setup and limited subsequent exploration of the
toolchain. Only one participant (P9) discovered mid-semester that Git
can be configured to sign commits automatically, switching away from
manually specifying the \texttt{-S} flag on each commit.

\subsubsection*{Key Storage Was Largely Passive and Security Was Inconsistent}

When asked how and where they stored their private key, participants ($n=14$) commonly reported relying on default storage locations provided by the tool, often without deliberate decision-making. P2 noted:

\longsay{I just left the key where it was generated and assumed that it was safe there.}{P2}

Similarly, P18 expressed uncertainty about the security of this default approach:

\longsay{anyone using my computer could easily access the key with Git and/or gpg commands.}{P18}

Several participants ($n=5$) were unable to recall the location of their private key at the time when completing their semester-end written reports:

\longsay{I forgot where I kept my private key. That is probably not a good thing.}{P7}

A smaller subset ($n=5$) described implementing additional security measures such as restrictive file permissions (\texttt{700}/\texttt{600}), full-disk encryption, or storing passphrases in external password managers. While $n=11$ participants reported using a passphrase, this was generally described as part of the setup process rather than a deliberate security decision. Only a small number of participants ($n=3$) demonstrated proactive key management beyond default configurations.

Taken together, these findings suggest that participants largely accepted default key management practices without active evaluation, with implications for secure-by-default design in cryptographic tooling.

\subsubsection*{Participants Linked Future Adoption Intent to Perceived Relevance}

Participants ($n=13$) expressed unconditional intent to continue signing commits after the course, with a subset ($n=5$) already doing so in repositories outside course assignments. Others ($n=7$) reported conditional intent, limiting future use to professional, collaborative, or high-stakes contexts. Only one participant ($n=1$) expressed no intent to continue, citing the absence of an audience for signed commits. In their written self-reports, participants most often explained their future intentions in terms of perceived contextual relevance rather than usability. Participants who emphasized security, authenticity, or provenance benefits ($n=9$), as well as those who highlighted professional or collaborative use cases ($n=8$), were predominantly among those expressing unconditional intent. In contrast, participants who described their repositories as personal, solo, or low-stakes ($n=5$) tended to report conditional or no future intent, regardless of usability ratings.

P14 illustrated this reasoning:
\longsay{I do not feel that my repos are `high stakes' enough for signed commits to be necessary. However, I'm glad that I now have experience in creating signed commits.}{P14}

Notably, participants with the lowest SUS scores (P7: 52.5; P11: 50.0) still reported conditional or continued intent, suggesting that usability alone did not determine adoption intent within this cohort.

\subsubsection*{Improvement Suggestions after Usage}
\label{sec:suggestions}

Participants offered improvement suggestions across five categories after using the system over the semester, with most concerns still relating to the setup process. The most common request was for a graphical interface, setup wizard, or tighter GitHub integration to replace manual configuration ($n = 6$). Examples included a GUI for the \texttt{allowedSignersFile} workflow (P2), a setup wizard integrated into the Git CLI (P4), and full GitHub Desktop integration (P14).

The second most frequent request was for consolidated and more complete documentation ($n = 7$). Participants described the current documentation as fragmented across authentication methods (GPG vs. SSH), inconsistent across platforms, and insufficiently covering the \texttt{allowedSignersFile} workflow. P20 expressed a preference for “a singular resource on GitHub that completely guides you through the process and any errors that may be encountered.”

Three additional categories each appeared in a small number of responses: more descriptive and actionable error messages ($n = 2$); tighter native integration of the signing workflow into Git, reducing reliance on external tools such as GPG and ssh-agent ($n = 2$); and automated or OS-level setup to remove the one-time configuration burden ($n = 2$).

\begin{mytakeaway}
Git commit signing becomes effortless once set up, and participants evaluated the semester-long experience positively. However, adoption did not necessarily lead to a deeper understanding of the underlying security model. Many participants were unaware of where their private key was stored, did not consider the implications of key loss or compromise, and did not intuitively sign every commit. Participants commonly explained their future signing intentions in terms of whether they perceived a personal need for it.
\end{mytakeaway}

\section{Findings: Cross-Device Setup}
\label{sec:second-device}

\subsection{Copy vs. New Key Comparison}
Participants ($n=22$) successfully signed and verified a Git commit from a second device, with an overall geometric-mean reported completion time of 51.04 minutes. Among the 20 participants with valid reported times at both time points, we observed no statistically significant difference between initial setup ($GM=44.18$ minutes; median $=52.5$; IQR $=37.5$) and second-device setup ($GM=51.99$ minutes; median $=42.5$; IQR $=78.75$; Wilcoxon $W=40.0$, $p=.255$, $r=.255$). Of those who completed the task, participants split between two credential strategies: transferring their existing private key ($n=12$) or generating a new keypair on the second device ($n=10$). The reasoning within each group was consistent.

Participants who generated a new key pair primarily cited security. Transferring a private key was understood as a practice to avoid, not merely a matter of preference:
\longsay{Private keys shouldn't be transferred between machines. Each machine should have its own private key. If the private key was copied, it would've existed in multiple places, increasing the risk of compromise.}{P8}
\longsay{I believe that sharing SSH keys among machines is probably not secure.}{P19}
Participants who copied their existing private key cited convenience and key consistency, specifically the desire to avoid registering a second public key with GitHub and to maintain a single verifiable identity across commits:
\longsay{Using the same key allows me to verify all commits together, rather than worrying about two keys.}{P4}
\longsay{I chose to copy my key so that I wouldn't have to add another public key to Github.}{P10}

\begin{table}
    \centering
    \caption{Comparison between participants who copied their existing private key to the second device and those who generated a new key pair.}
    \label{tab:ktd-comparison}
    \small
    \setlength{\tabcolsep}{3pt}
    \resizebox{\columnwidth}{!}{%
    \begin{tabular}{lcccccc}
    \toprule
    \textbf{Metric} &
    \textbf{Overall ($n=22$)} &
    \textbf{Copy ($n=12$)} &
    \textbf{New ($n=10$)} &
    \textbf{$U$} &
    \textbf{$p$} &
    \textbf{$r$} \\ 
    \midrule
    ASQ, $M$ ($SD$)
    & 5.32 (1.51)
    & 5.28 (1.77)
    & 5.37 (1.20)
    & 58.0 & .921 & .021 \\ 
    Time (min)$^a$
    & 51.04 [42.50; 67.50]
    & 51.33 [52.50; 78.75]
    & 50.69 [40.00; 30.00]
    & 57.0 & .868 & .036 \\ 
    \bottomrule
    \end{tabular}%
    }
    \smallskip

    \raggedright\footnotesize
    $^a$ Time cells report geometric mean [median; IQR].
    All comparisons used two-sided Mann--Whitney $U$ tests and
    are exploratory.
\end{table}

Neither approach was universally smoother. We found that participants who generated a new key pair avoided transfer friction entirely but faced the same initial configuration steps as in the first setup (\S\ref{sec:results-first-setup}). Participants who copied their existing key gained a familiar credential but introduced new friction around export, transfer, and import. Importantly, despite the different forms of friction, the two paths produced similar outcomes: the exploratory comparisons did not detect statistically significant differences in satisfaction or completion time, and satisfaction remained positive in both groups (Table~\ref{tab:ktd-comparison}). The following subsections detail the specific steps and challenges participants encountered along each path.

\subsection{Reconfiguration and Key Transfer Friction}
\label{subsec:p2-friction}

Participants ($n=14$) encountered at least one Git or signing
configuration error during the second-device setup. Friction clustered
around two sources: key transfer mechanics for participants who copied
their previous key, and persistent Git reconfiguration gaps affecting
both groups.

\subsubsection{Reconfiguration on the New Machine}
\label{subsubsec:reconfig}

The most prevalent friction source was not key transfer itself but
reconfiguration from scratch on the new machine: six participants
(P2, P5, P9, P10, P11, P20) encountered the \texttt{user.signingkey}
omission or \texttt{allowedSignersFile} schema errors described in
\S\ref{subsubsec:allowed-signers}, confirming both as persistent
gaps rather than one-time setup failures.

\subsubsection{Key Export and Import}
\label{subsubsec:kxf}

Among participants who copied their existing key ($n=12$), the majority
($n=9$; P3, P4, P10, P12, P13, P16, P17, P18, P21) needed to look up
the correct export flags, as GitHub's main signing documentation does
not cover key transfer between machines. The
\texttt{--export-secret-keys} flag in particular was not discoverable
from the primary guide. P13 noted that GitHub's documentation covered
exporting the \emph{public} key but \enquote{the private key \dots\ is
not listed,} requiring a separate search. P21 similarly consulted three
separate external resources to complete the export and import steps,
none of which were reachable from GitHub's primary signing guide. Most
of these participants ($n=8$; P3, P4, P10, P13, P16, P17, P18, P21)
similarly needed external sources to complete the import step on the
destination machine, a workflow also absent from GitHub's documentation.

Once located, the mechanics were unremarkable: P3 described the export as having \enquote{worked flawlessly} once the commands were known, and P16 found the transfer trivial: \enquote{two commands to make the files,
and MobaXterm allowed me to drag and drop them.}

\subsubsection{Key Trust for Imported Keys}
\label{subsubsec:trust}

One friction point was unique to participants who copied their existing
key: GPG requires imported keys to be explicitly trusted before they
can be used without a warning. This requirement does not apply to
locally generated keys and is absent from GitHub's main signing guides.
P17 encountered it unexpectedly after what they believed was a
successful setup:

\longsay{After signing, I went to verify my signature when I saw a
message that read ``WARNING: This key is not certified with a trusted
signature!'' \dots\ I had to trust the public key. This was not needed
before because the keys were generated on my computer, but imported
keys must be trusted.}{P17}

P21 encountered the same requirement and documented it explicitly,
working through the \texttt{gpg --edit-key} trust dialogue and setting
the trust level to ultimate before the key could be used without a
warning. The warning appears at verification time rather than at import,
meaning participants who skipped manual verification would not have
encountered it until a downstream consumer of the signed commit flagged
the untrusted key. This mirrors the underdocumented dependencies
identified in \S\ref{sec:results-first-setup}.

\subsubsection{Key File Location}
\label{subsubsec:locate}

P13 spent considerable time locating their existing key files prior to
transfer, encountering a shell-specific discrepancy:
\texttt{gpg --list-secret-keys} returned nothing in PowerShell but
succeeded in Git Bash --- the shell originally used to create the key.
No error message explained the discrepancy; the command simply returned
an empty keyring. This surface-level inconsistency added approximately
30 minutes of unproductive debugging before the transfer could begin.

\begin{mytakeaway}
Extending commit signing to a second device is manageable but does not
get easier. The toolchain has no memory across machines, so users who
want to manage their own keys rather than delegate that responsibility
to the platform must bear the full cost of reconfiguration every time
they add a new device.
\end{mytakeaway}

\section{Findings: Identifying Malicious Commits}
\label{sec:finding-bad-commit}

\subsection{Detection Performance}

Table~\ref{tab:commit_categories} shows detection rates, defined as the proportion of participants who identified at least one commit in each category. The most commonly identified issues were unknown developer commits and commits signed with the wrong key. Timestamp manipulation was the least frequently detected category ($n = 9$). Participants flagged commits dated in 2015 and 2052. Notably, several participants who identified the 2052 future-dated commit missed the 2015 backdated commit, suggesting that anomalies in the past are harder to detect than those in the future.

\begin{table}
    \centering
    \caption{Categories of Problematic Commits and Detection Rates}
    \label{tab:commit_categories}
    \setlength{\tabcolsep}{4pt}
    \small
    \begin{tabular}{lrr}
    \toprule
    Category & \#Commits & Detected \\
    \midrule
    Unknown developer                  & 3 & 12/22 (55\%) \\
    Timestamp manipulation             & 2 &  9/22 (41\%) \\
    Author/committer mismatch          & 2 &  9/22 (41\%) \\
    Wrong signing key                  & 1 & 11/22 (50\%) \\
    Missing signatures                 & 5 & 11/22 (50\%) \\
    \bottomrule
    \end{tabular}
\end{table}

Overall performance was mixed. About a quarter of participants ($n=6$, \pct{6}) failed to identify any anomalous commits. Although participants approached the task with suspicion, none identified all 13 problematic commits. (See Appendix~\ref{sec:supplementary-results}, Table~\ref{tab:score_distribution})

We excluded three participants (P3, P10, P18) from time-based analyses due to missing time reports. Among the remaining participants, those who spent more time on the task identified more anomalous commits (Pearson $r = 0.668, p = 0.002;$ Spearman $\rho = 0.631, p = 0.004$). In contrast, self-reported satisfaction was not associated with performance (Pearson $r = -0.081, p = 0.726;$ Spearman $\rho = -0.137, p = 0.553, n = 21$), indicating that perceived ease of the identification process did not predict detection accuracy. Finally, to assess whether familiarity with the signing mechanism influenced detection performance, we compared GPG users ($n = 16$) and SSH users ($n = 6$). Despite differences in mechanism, the exploratory comparison did not detect a statistically significant difference in detection performance between GPG and SSH users (mean: $5.06$ vs.\ $4.00$; median: $3.5$ for both; Mann--Whitney $U = 55.0$, $p = 0.6268$; Appendix~\ref{sec:supplementary-results}, Table~\ref{tab:inferential-verification}). Verification difficulties were therefore observed in both groups, although this result does not establish equivalent performance.

\subsection{Verification Process and Barriers}

Participants used three main strategies: the Git CLI alone ($n=10$),
the GitHub web interface alone ($n=2$), or a combination of both
($n=9$). One participant (P18) did not attempt the task. The choice of strategy had little bearing on the number of commits correctly flagged (CLI mean: 5.00; CLI+web mean: 5.11; web mean: 4.50). Instead, the largest observed group difference involved whether a participant successfully configured the \texttt{gpg.ssh.allowedSignersFile} for SSH signature
verification, since we provided two SSH keys for the authorized
contributors. Without this file correctly configured, \texttt{git
log --show-signature} cannot match commit signatures against the
provided keys, leaving participants unable to distinguish a properly
signed commit from an unsigned one --- regardless of whether they used
the CLI or the web interface. Participants who configured it correctly scored significantly higher on anomaly detection than those who did not (Appendix~\ref{sec:supplementary-results}, Table~\ref{tab:inferential-verification}).


\subsubsection{allowedSignersFile Configuration Failures}

About half of the participants ($n=12$, \pct{12}) attempted to configure the \texttt{allowedSignersFile}; of these, there was a 33\% failure rate (4/12) among those who actively engaged with the tool. Across the full cohort ($n=22$), participants ($n=7$)—including both those who eventually succeeded and those who failed—confused the provided SHA256 fingerprints with full SSH public keys. This confusion stemmed from a lack of awareness that they needed to retrieve full public keys from GitHub’s \texttt{.keys} endpoint separately. P3 described the experience:

\longsay{This activity took MUCH longer than it should have. During the process, I went down a rabbit hole, thinking that I needed to add these keys to a verified signer list \ldots\ This was an extreme waste of
time.}{P3}

P15 captured the underlying confusion directly:

\longsay{I knew that the hashes were key fingerprints, but otherwise I
was not sure what to do with them.}{P15}

P10 encountered the same wall, noting that \longsay{all the information
sources I found assumed one already had the public key.}{P10} P4
similarly tried to convert the fingerprints into usable keys using
\texttt{ssh-keygen}, but abandoned the attempt after concluding that
\longsay{these keys are not keys, at least in any format I know
of.}{P4}

P13 came closest, eventually finding the full public keys, but then
spent three hours debugging CRLF line-ending mismatches in the signers
file before an AI assistant identified the formatting issue, reflecting
that \longsay{no single source of information I found showed how to do
this correctly, and it took multiple trial-and-error efforts to
configure it appropriately. Even as it is, adding signatures to this
file is a painful, manual experience that I would not enjoy repeating
for multiple users.}{P13}

We also found that participants failed to reach a successful configuration ($n=14$). Also, some participants explicitly gave up on cryptographic verification entirely ($n=6$). Some ($n=4$) abandoned the task after failing to reach a successful configuration, while two ($n=2$) successfully configured the \texttt{allowedSignersFile} but gave up due to confusion between SSH and GPG mechanisms. They instead relied on surface-level cues instead of verifying signatures. This ``gave up'' group achieved a mean score of 1.5 (out of 13), substantially lower than those who persisted ($M=6.00$; Mann-Whitney $U=18.0$, $p=.027$, $r=.470$). P4 summarized the experience:

\longsay{I would have liked more guidance in how to verify commits using
another developer's keys, because it was difficult to figure out.
Ultimately, with all the warning signs I saw, I determined the
repository was not safe, but I ended up never using the keys.}{P4} P8
reached a similar dead end, concluding incorrectly that no commits were
signed at all: \longsay{I realized that the assignment only provided a
SHA256 fingerprint, not the full SSH public key. Since the commits
weren't signed, verifying them wouldn't work anyway.}{P8} P10 attempted
to download the developers' public keys directly but abandoned the
effort, remarking that \longsay{the process is fine if you just use the
GitHub website which will show when commits are properly signed. Doing
it in the CLI is a waste of time.}{P10}

A distinct failure mode was encountered by P22, whose attempt to run
\texttt{git log --show-signature --decorate --oneline} produced a core
dump, causing Git to crash entirely. Rather than treating this as a
tooling error, P22 interpreted the crash as evidence of repository
corruption and abandoned signature-based verification entirely. This is
a meaningful edge case: a tool failure was misread as a security signal,
which both inflated P22's perceived suspicion of the repository and
prevented any further cryptographic analysis.

It is worth noting that the \texttt{allowedSignersFile} configuration
was not strictly necessary to complete this task. The \texttt{git log
--pretty=format} command exposes signature status (\texttt{\%G?}), the
signing key fingerprint (\texttt{\%GK}), and the signer identity
(\texttt{\%GS}) directly, allowing participants to manually compare
fingerprints against the two provided keys without any local key
configuration. Furthermore, providing fingerprints rather than full
public keys reflects a plausible real-world scenario --- a maintainer
may share only a fingerprint for verification purposes, and in practice
it is the responsibility of the consuming developer to verify and
maintain their own security posture independently. The fact that twelve
participants attempted --- and several failed --- to configure the
\texttt{allowedSignersFile} suggests they were unaware that a simpler
path existed, pointing to a discoverability problem in Git's signature
verification tooling rather than a gap in participant ability. P3
observed that online resources were \longsay{incredibly unhelpful
because most of them said that you could add allowed signers keys but
often did not explain the file format to do so. If they had the format,
they did not show what the key should look like or include any fake
examples of a file.}{P3}

\subsubsection{GPG--SSH Mechanism Confusion}

A distinct but related barrier was confusion between GPG and SSH
signing mechanisms. Several participants arrived at this task with prior
experience of GPG signing from earlier in the course, and attempted to
apply the same approach here --- only to discover that the repository
used SSH signatures instead. P8 described this as the hardest part of
the activity: \longsay{the error messages were confusing because I had
only worked with GPG signatures and not SSH signatures, plus the errors
didn't really explain what was wrong or provide information on how to
fix it --- it required me to understand multiple signature
mechanisms.}{P8} P10 attempted to download the developers' GPG keys
using \texttt{wget} and \texttt{gpg} before realizing they were
irrelevant, noting that \longsay{the only one I could get was expired
anyway.}{P10} P11 similarly attempted GPG verification before
abandoning it, observing that \longsay{there was only one person who
had a GPG key, and it just did not have any usage for signatures
here.}{P11}

This confusion was compounded by Git supporting multiple signing
mechanisms --- GPG, SSH, and S/MIME --- with no unified verification
workflow and error messages that did not clarify which was in use.

\subsubsection{Code Content vs.\ Signing-Based Analysis}

Five participants (P1, P6, P7, P14, P16) augmented their signing
analysis by manually reviewing commit code content for malicious
patterns. P1 used \texttt{git log} with grep to search for suspicious
keywords, noting: \longsay{I used git log with grep to look for API,
password, key, token, eval, exec, decode, etc.\ phrases within the code
files. I found a hit in one commit that leaked the API key and had an
eval() function.}{P1} P16 opened each commit individually on GitHub,
describing the process as \longsay{skimming the contents of the commit
for anything that looked malicious.}{P16}

This approach produced similar descriptive detection rates (content reviewers: mean $= 4.60$; non-reviewers: mean $= 4.82$), and in one case led to an incorrect conclusion. P14 reviewed all nine unverified commits manually and concluded:

\longsay{I was able to manually read through all of the code in these
commits and determine that none of them contain anything that appears to
be malicious. Therefore, I do not think that any of the commits in this
repository are genuinely problematic.}{P14}

A commit may contain benign code and still originate from an
unauthorized party --- the purpose of commit signing is to verify
\textit{who} made a change, not whether the change itself is harmful.

\begin{mytakeaway}
Commit signing is only useful if someone can verify the commits. We
found that for our participants, verification is too hard to do correctly. A simpler verification path existed that required no additional
configuration, but most participants did not discover or understand it.
Those who tried the harder path mostly failed. Those who gave up relied
on surface-level cues that commit signing is specifically designed to
replace. If verification is too hard to do correctly, the signing
itself provides a false sense of security.
\end{mytakeaway} 
\section{Findings: Security Understanding}
\label{sec:understanding}

Table~\ref{tab:thought_per_participant} (Appendix~\ref{sec:supplementary-results}) reports participant-level classifications across the five questions, while Table~\ref{tab:thought_aggregate} summarizes aggregate classifications by question. Despite their advanced technical training, nearly half of participants ($n = 10$, \pct{10}) provided at least one response classified as a misconception or wrong: seven had misconceptions but no wrong responses, whereas three (P10, P15, and P16) had at least one wrong response. We describe the recurring misconceptions below.

\begin{table}
    \centering
    \caption{Aggregate correctness by open-ended security-reasoning prompt. 
    C = Correct, M = Misconception, W = Wrong. Denominators vary 
    by question because some participants did not answer all questions; see Table~\ref{tab:thought_per_participant} 
    for individual responses.}
    \label{tab:thought_aggregate}
    \small
    \setlength{\tabcolsep}{4pt}
    \begin{tabular}{lrrrr}
    \toprule
    Question & $n$ & C & M & W \\
    \midrule
    Q1: Benefits/drawbacks  & 21 & 17 (81\%) & 3 (14\%) & 1 (5\%)  \\
    Q2: Single vs.\ new key & 20 & 19 (95\%) & 1 (5\%)  & 0 (0\%)  \\
    Q3: Lost key            & 20 & 13 (65\%) & 6 (30\%) & 1 (5\%)  \\
    Q4: Stolen key          & 18 & 14 (78\%) & 2 (11\%) & 2 (11\%) \\
    Q5: Attack vectors      & 20 & 20 (100\%)& 0 (0\%)  & 0 (0\%)  \\
    \bottomrule
    \end{tabular}
\end{table}

    \subsubsection*{Key Loss Invalidates Existing Signatures}

    The most common misconception ($n=3$; P3, P9, P11) was that losing a private key makes previously signed commits unverifiable. P3 wrote that: \longsay{your old commits may not be verifiable as your commits anymore and you cannot trust them}{P3} P9 reached the same conclusion, and P11 dismissed the concern too quickly, overlooking the exposure window between loss and revocation entirely. This is incorrect. A commit signature is stored permanently in the repository at the time of signing. Verifying it requires only the public key, which remains available regardless of whether the private key is later lost. This misconception confuses the private key's role in \textit{signing} with its role in \textit{verifying}, which are deliberately independent operations in asymmetric cryptography.


    \longsay{I don't think there are any security issues once the new key is instated and the old one revoked}{P11}

    \subsubsection*{SSH Push Keys Already Authenticate Commits}
    
    P10 argued that commit signing was \longsay{mostly just a 
    waste of time — you already need an SSH key to push commits 
    to GitHub so no one else should be able to impersonate 
    you}{P10} This confuses SSH push authentication, which 
    authenticates the \textit{connection} to the platform, with 
    commit signing, which authenticates the \textit{content} of 
    the commit. Git commit metadata is trivially spoofable 
    without signing regardless of who holds the SSH push key~\cite{holtgrave2025attributing}. A developer with this belief would see no reason to adopt commit signing at all.
    
    \subsubsection*{Account Actions Substitute 
    for Key Revocation}
    
    Participants ($n=2$; P15, P16) proposed account-level 
    responses to key theft that do not address the cryptographic 
    problem. P15 suggested moving everything to a new repository 
    and P16 proposed creating a new GitHub account. Neither 
    action revokes the compromised key or prevents an attacker 
    from continuing to sign commits with it. These responses 
    suggest that some participants defaulted to 
    \textit{account} lifecycle thinking rather than 
    \textit{cryptographic key} lifecycle thinking. Revocation 
    must happen at the key level to be effective.

    \subsubsection*{Usability Risk as Argument Against Signing}

    Participants ($n=2$; P7, P16) argued that the risks of poor key
    management justified not signing commits at all. P16 argued that
    forcing non-technical users to sign could \longsay{compromise their
    keys, which is more of an issue than just having unsigned
    commits}{P16} Both responses confuse the \textit{risk of poor
    implementation} with the \textit{security value of the mechanism}.
    Poor implementation is a usability problem; unsigned commits are a
    security problem. The solution is better tooling, not abandoning
    signing. P7 also held a secondary misconception, suggesting that a
    finder of the lost private key \longsay{could presumably easily
    generate the public key and start sending off signed commits}{P7} ---
    misunderstanding that the public key is already public by design.

    \subsubsection*{Signing Key Reuse Is Bad 
    Practice}
    
    P15 incorrectly applied a prior course lesson: \longsay{we 
    have been told repeatedly that it is bad to reuse private 
    keys, so this might apply to signing keys as well}{P15} The 
    do-not-reuse-keys rule applies to session and encryption 
    keys, where reuse enables cryptanalytic attacks. Signing keys 
    are designed for repeated use across many commits and reuse 
    is standard practice. This illustrates a broader risk in 
    security education: rules learned in one context can be 
    incorrectly transferred to another.

\begin{mytakeaway} Nearly half of participants held at least one misconception about commit signing security, despite a semester of correct procedural use. Several responses reflected incorrect transfer from other security contexts: participants applied rules that do not apply to commit signing, proposed account-level responses when key-level actions were required, or treated another authentication mechanism as a substitute for signing. Thus, procedural fluency was not consistently accompanied by correct reasoning about commit-signing security properties.

\end{mytakeaway}


\section{Study Limitations}
\label{subsec:limitation}

First, the commit verification task used SSH-signed commits, whereas many participants had used GPG based signing earlier in the semester. Including both mechanisms may have increased ecological validity, but switching mechanisms may also have introduced additional difficulty. As reported in \S\ref{sec:finding-bad-commit}, we observed no statistically significant performance difference between participants who had used GPG and those who had used SSH. Furthermore, GitHub recommends SSH as the simpler and more accessible signing mechanism~\cite{github_commit_signing}, suggesting that developers may increasingly encounter SSH-signed commits regardless of which mechanism they initially learned.

Second, the verification task contained a higher anomaly density than developers would typically encounter, with 13 anomalous commits among 20, and participants received key fingerprints rather than a pre-configured list of trusted signers. These choices may not represent every real-world verification scenario. They were deliberate: the anomaly density provided sufficient behavioral observations within one session, while fingerprint comparison represented the high-stakes task of verifying unfamiliar contributors. Nevertheless, these choices may have affected task difficulty. Participants' difficulties with legitimate commits, reported in \S\ref{sec:finding-bad-commit}, indicate that the observed problems were not attributable solely to anomaly density.

Third, we acknowledge that the three-month study period captured sustained signing across multiple submissions but not long-term key-lifecycle management, including key expiration, rotation, revocation, and recovery after loss or compromise. However, as in \S\ref{sec:understanding}, we examine participants' reasoning about key loss, theft, and revocation. Still, we note that our findings should not be generalized to developers' performance during long-term key-lifecycle events.

Fourth, while we scope our single-university opt-in sample as a proxy for junior developers encountering unfamiliar security workflows (\S\ref{subsec:participants}), this limits external validity regarding professional incentives and organizational support. Future work must test whether our observed security-understanding gaps and verification failures persist among industry practitioners, particularly within mature software supply chain projects.

Fifth, as discussed in \S\ref{subsec:relevance-and-voluntary-use}, the required-course setting limits our findings to sustained use under an external requirement rather than voluntary adoption. Furthermore, a graded environment may introduce demand characteristics, prompting participants to report more positive experiences or take unrepresentative care. We partially mitigated this by grading self-reports solely for completion, emphasizing that security prompts lacked single correct answers, and corroborating qualitative claims with behavioral \texttt{.git} logs; however, this limitation cannot be fully eliminated. Finally, as detailed in \S\ref{subsec:procedures}, because we did not track post-semester Canvas activity, we cannot confirm whether participants re-examined their read only submissions before consenting to research use.
\section{Discussion and Conclusion}
\label{sec:discussion}

In this paper, we evaluated how 22 computer science students, representing junior developers, experienced adopting and sustaining Git commit signing. We discuss our findings in the following subsections.

\subsubsection*{Beyond Simple Adoption}

A key takeaway from our work is that we cannot measure repository security just by counting signed commits in repository histories~\cite{shittu2026analysis,holtgrave2025attributing,10.1145/3756681.3756959,kalu2025industry}. While checking Git logs proves that developers are following the steps to sign their code, it hides a major blind spot: it does not show whether developers actually know how to verify signatures, understand what verification results mean, or handle key-management exceptions. Within this cohort, our findings show a gap between following these routine steps and demonstrating the security understanding needed for effective use. Even though our participants easily made commit signing a daily habit and liked using it, they remained vulnerable when faced with actual security tasks. During the verification task, many participants missed anomalous commits, and several demonstrated misunderstandings about what happens when keys are lost, stolen, or revoked. This pattern suggests that, within this cohort, learning how to generate signed commits did not automatically produce the understanding needed to manage keys securely or identify repository threats. Because tooling difficulties and gaps in security understanding appeared together, reducing setup friction alone may be insufficient to address the problems observed in this cohort. We therefore argue that commit signing must be treated as a complete three-part workflow: signing, verification, and exception handling. To be truly secure, projects cannot just generate signatures; they must also check them against authorized identities and have clear plans for dealing with missing signatures, unknown keys, and stolen credentials. Ultimately, this shifts the focus for the entire software community: generating signed commits is simply the starting point of a much broader security process.

\subsubsection*{Barriers to Effective Verification}

We do not interpret our results as evidence that tooling, education, or security understanding is the single underlying barrier. Instead, our verification findings illustrate how technical friction and conceptual ambiguity interact to derail the verification process. On the technical side, the current Git architecture imposes heavy manual configuration burdens before a developer can even attempt signature verification. By requiring out-of-band public key discovery, manual formatting of the \texttt{allowedSignersFile}, and navigation across disparate signing standards (GPG versus SSH), existing tools treat verification as an advanced, optional task rather than a core security primitive. Because successful configuration in our study was the primary prerequisite for anomaly detection, poor tool discoverability and opaque error messaging directly inhibit effective threat detection.

However, removing configuration hurdles would not necessarily address all of the conceptual misunderstandings observed during these tasks. When developers attempt to evaluate repository trust by manually inspecting code syntax for malware, or when they treat platform push authentication as a substitute for commit signing, they reveal a fundamental conflation of access control, signature validity, signer authorization, and code safety. As highlighted in literature~\cite{holtgrave2025attributing,zhang2025pushing}, cryptographic validity alone offers negligible protection against impersonation attacks if signatures are not evaluated against authorized identities. Yet in daily practice, many developers default to relying on platform-level verification indicators without understanding this exact boundary—specifically, assuming an indicator proves developer authorization when it may only confirm cryptographic validity.

Because technical friction and conceptual misunderstandings interacted during the task, our findings suggest that verification failures are unlikely to be addressed by tooling or education in isolation. Addressing both barriers likely requires tools that make identity binding and signature verification more discoverable and usable, together with education that explains the security guarantees and limitations of verification.

\subsubsection*{Relevance and Voluntary Use}
\label{subsec:relevance-and-voluntary-use}

A major takeaway from this paper is that ease of use alone may be insufficient to support voluntary adoption when developers do not perceive a practical reason to use the mechanism. While participants initially used commit signing because it was required for the course, their end-of-semester written self-reports more often explained future signing intentions in terms of perceived usefulness than usability. Specifically, participants who recognized a security threat or collaborative benefit were more likely to express an intention to continue signing, including some who experienced setup difficulties. Participants who described their repositories as personal, solo, or low-stakes were more likely to question the practical value of commit signing because they did not expect others to verify their signatures. In contrast, participants who emphasized provenance or multi-developer collaboration were more likely to express an intention to continue signing, including when they reported tooling frustration.
This behavior aligns with prior usable-security research showing that users weigh perceived benefits against daily effort before adopting security habits~\cite{fagan2016motivates}. Commit signing may appear less relevant in a private repository where the author is the only contributor, while its potential value may be more apparent in collaborative projects that must address impersonation threats. Therefore, efforts to increase adoption across the software industry cannot focus solely on improving user interfaces. Future tools, training, and research must focus heavily on communicating the threat model so developers actually understand why signing matters to their daily workflow.

\subsubsection*{Supporting the Complete Workflow}

To build a truly resilient software supply chain, the security community must stop treating tool usability and developer education as competing alternatives; securing the commit-signing workflow requires treating both as two sides of the same coin. First, tooling must remove the setup and documentation friction that currently prevents developers from completing verification. Interfaces must also clearly communicate what verification results actually mean by distinguishing among four distinct security states: a missing signature, a cryptographically invalid signature, a valid signature from an unknown key, and a valid signature from a known but unauthorized key. Because each state carries a different level of risk, tools must make these differences obvious so developers can respond appropriately.

Second, education must move beyond simply teaching developers how to generate a key and turn on signing. Instruction needs to explain how projects bind keys to authorized identities, why logging into a repository is fundamentally different from signing a commit, and what a valid signature actually proves. Training must also prepare developers for messy, real-world exceptions, including how to handle unknown signers, key loss, credential theft, key rotation, and revocation.

Finally, tooling and education must actively reinforce each other. As prior usable-security research demonstrates, tools should display explanations right when a developer faces a security decision, rather than expecting them to memorize external documentation~\cite{egelman2008warned}. Education then provides the underlying threat model needed to interpret those explanations. This synergy is especially important for platform-assisted verification, such as GitHub's verification indicators. While platform support reduces local setup burdens, interfaces must make the underlying trust decision explicit. Developers should be able to tell at a glance whether the platform only verified the cryptographic math, or if it also confirmed that the signing key was authorized for that specific identity and repository. Prior work proves that web-based repository tools can automate these checks while remaining compatible with Git signatures~\cite{afzali2018legitimate}. Ultimately, the goal is to cut unnecessary developer effort without sacrificing clear, auditable security guarantees.

\subsubsection*{Conclusion}

Taken together, our findings shift the goal from maximizing the number of signed commits to supporting correct trust decisions across signing, verification, and exception handling. Achieving that goal requires usable tools, education tied to practical security decisions, and development contexts in which users understand the relevance of signing.

\begin{acks}
The authors retained full responsibility for all content in this paper. 
Generative AI tools (Claude Sonnet 4.5~\cite{anthropic2026claude}, 
Grammarly~\cite{grammarly2026}, and Gemini 3~\cite{google2026gemini}) 
were employed exclusively for light copy-editing tasks such as correcting 
grammar, fixing typographical errors, and improving sentence clarity. 
No substantive content was produced or modified by these tools.
\end{acks}

\bibliographystyle{ACM-Reference-Format}
\bibliography{ref}

\appendix

\section{Open Science}
\label{sec:open-science}

All artifacts for this paper are available at the following anonymous repository:
\url{https://anonymous.4open.science/r/git-commit-signing-longitudinal-user-study-artifact/README.md}

We release the assignment instructions for both parts of the study, our analysis codebook, redacted consent and recruitment forms, and the full analysis scripts used to process participant repositories and conduct statistical tests. On the other hand, participant Git repositories and public keys cannot be released as they contain personally identifiable information, including participant names, email addresses, and
institutional affiliations embedded in Git commit metadata. Similarly, the raw data collected from participants, which includes free-text qualitative responses, cannot be released as it constitutes human subjects data collected under IRB oversight and could be used to identify individual participants. Releasing either of these artifacts would violate the privacy of study participants and the terms under which IRB approval was granted.

\section{Ethical Considerations}
\label{ethics}

\subsection{Research Procedures}
\label{subsec:procedures}

The study was IRB-approved and embedded in a required applied cryptography 
course. The \texttt{.git} submissions, public keys, written self-reports, and SUS and ASQ responses were completed as graded course components by enrolled 
students before consent was requested. For students who did not subsequently 
consent, these materials remained course records and were not extracted, 
coded, analyzed, or reported as research data. Research participation was 
opt-in ($n{=}22$ of $70$), and withdrawal carried no academic consequence.
Participants were not compensated, as the assignments were a required part 
of the course.

To preserve academic integrity and avoid any potential influence on student 
performance, we did not disclose the research aims during the semester, nor did we begin data analysis until after the semester had concluded and final grades had been released. At that point, we contacted students via email to solicit consent. The email explained that the goal of the research was not to evaluate the students themselves, but rather to assess the state of Git commit signing as a usable key management system, and that all data would be anonymized prior to analysis. Students were directed to a consent form where they could formally agree to participate. In the consent form, we further informed participants that their decision to participate or withdraw would have no bearing on their grades or standing, and that withdrawal was possible until analysis of the anonymized 
data began.

Students retained read-only access to their assignments in Canvas after 
completing the course. They could view their previous submissions but could 
not modify them. Because consent was obtained after the course had concluded 
and final grades had been released, students could not change their 
submissions after learning about the study but before deciding whether to 
participate. We did not record whether students viewed their submissions 
before providing consent. Only submissions from consenting students were 
included in the research dataset and analyzed.

The course was also designed with participant welfare in mind. Because 
generating and managing cryptographic keys for the first time can be 
unfamiliar and time-consuming, and students had competing coursework 
obligations, we embedded structured support into the assignment. Students 
who remained stuck after 90 minutes were encouraged to contact the 
teaching assistants for guidance, and after three hours of effort, direct 
assistance was provided to ensure no student was blocked from completing 
the course requirements. Any assistance received was noted in the 
student's report but did not affect eligibility to receive full credit. 
Overall, this study prioritized the students' educational goals over the 
research goals.

\subsection{Stakeholders}

Direct stakeholders are the research team, who bear reputational risk if 
privacy protections fail, and the 22 consenting participants, whose burden 
is minimal as their work was already completed. Indirect stakeholders 
include Git, GnuPG, and GitHub developers, whose tool designs are 
implicated in our findings; the open-source community, who stand to 
benefit from improved signing practices; and potential malicious actors, 
who represent a dual-use risk addressed below.

\subsection{Menlo Principles}

\subsubsection*{Respect for Persons.} Autonomy is protected by opt-in consent, 
completion-only grading, and post-grade consent solicitation. All 
participants are anonymized as P1--P22 and findings are reported at the 
population level only.

\subsubsection*{Beneficence.} Participant risk is low: no burden beyond 
already-completed coursework was imposed, and \texttt{.git} log data 
provides a behavioral cross-check on self-report findings. The expected 
benefit---the first multi-month evidence of the gap between procedural 
adoption and security understanding in commit signing---flows to the 
broader developer ecosystem.

\subsubsection*{Justice.} Recruitment was determined by course enrollment. 
Although the burden of participation falls on students while benefits flow 
to the broader community, we judge this asymmetry acceptable because 
participation required no additional effort and was entirely voluntary.

\subsubsection*{Respect for Law and Public Interest.} The study complies with 
IRB requirements and GitHub's Terms of Service. Data are stored on a 
secured institutional server restricted to the research team. Identifying 
information will be deleted upon study completion; any shared data will be 
fully de-identified. Methods and findings are reported transparently to 
support replication.

\subsection{Harms and Mitigations}

\subsubsection*{Power imbalance (tangible harm).} Students may moderate responses 
to please the instructor. Mitigated by completion-only grading, explicit 
framing of open-ended questions as having no correct answer, post-grade 
consent solicitation, and \texttt{.git} log cross-checks. The residual 
risk of self-censorship in written self-reports cannot be fully eliminated 
and is acknowledged as an unmitigated limitation.

\subsubsection*{Privacy (rights violation).} Written self-reports and signing patterns could expose individuals to reputational harm. Mitigated by full anonymization, aggregate-level reporting, exclusion of contextually 
identifying quotes, and deletion of all identifying information upon study 
completion. Residual risk is low.

\subsubsection*{Dual-use (third-party tangible harm).} Identified failure modes 
could inform adversarial targeting. Mitigated by reporting findings at the 
conceptual level only, omitting operational details, and framing all 
results as defensive interventions.

\subsection{Decision to Conduct and Publish}

The decision to conduct the research is supported. 
Consequentially, the burden on participants is minimal, residual harms are 
low after mitigation, and the benefit to ecosystem security yields strongly 
positive net utility. Deontologically, no participant's right to informed 
consent is violated, no participant is used solely as a means, and privacy 
rights are protected throughout. Both frameworks likewise support 
publication. Consequentially, withholding findings would allow a false 
assumption of security coverage to persist in supply-chain frameworks, 
imposing an ongoing ecosystem cost that outweighs the residual dual-use 
risk. Deontologically, the security community and downstream software 
consumers have a legitimate claim to accurate information about the limits 
of a mechanism on which they rely. We report limitations and context 
throughout and present all findings at the population level to protect 
individual participants.

\section{Supplementary Results}
\label{sec:supplementary-results}

\begin{table*}[t]
\centering

\begin{minipage}[t]{0.47\textwidth}
\vspace{0pt}
\centering

\captionof{table}{Participant demographics ($N=22$).}
\label{tab:demographics}

\footnotesize
\setlength{\tabcolsep}{5pt}
\renewcommand{\arraystretch}{0.85}

\begin{tabular}{lll}
\toprule
Participant & Gender & Level \\
\midrule
P1  & Male              & Senior     \\
P2  & Female            & Senior     \\
P3  & Female            & Senior     \\
P4  & Male              & MS student \\
P5  & Female            & Senior     \\
P6  & Female            & Senior     \\
P7  & Male              & Senior     \\
P8  & Female            & MS student \\
P9  & Male              & Senior     \\
P10 & Male              & MS student \\
P11 & Male              & Senior     \\
P12 & Prefer not to say & Prefer not to say \\
P13 & Male              & Senior     \\
P14 & Male              & Senior     \\
P15 & Male              & MS student \\
P16 & Male              & MS student \\
P17 & Female            & Senior     \\
P18 & Male              & Senior     \\
P19 & Male              & Senior     \\
P20 & Male              & Senior     \\
P21 & Male              & MS student \\
P22 & Male              & Senior     \\
\bottomrule
\end{tabular}

\vspace{1.5cm}

\captionof{table}{Per-participant coding across five open-ended
security-reasoning prompts. C = Correct, M = Misconception,
W = Wrong, --- = not answered. $n$ varies due to incomplete
responses (P2, P10, P18, P22).}
\label{tab:thought_per_participant}

\footnotesize
\setlength{\tabcolsep}{3pt}
\renewcommand{\arraystretch}{0.85}

\begin{tabular}{lccccc}
\toprule
    & Q1 & Q2 & Q3 & Q4 & Q5 \\
\midrule
P1  & C   & C   & C   & C   & C   \\
P2  & C   & --- & --- & --- & --- \\
P3  & C   & C   & M   & C   & C   \\
P4  & C   & C   & C   & C   & C   \\
P5  & C   & C   & C   & M   & C   \\
P6  & C   & C   & M   & C   & C   \\
P7  & M   & C   & M   & C   & C   \\
P8  & C   & C   & C   & C   & C   \\
P9  & C   & C   & M   & C   & C   \\
P10 & W   & C   & M   & --- & C   \\
P11 & C   & C   & M   & C   & C   \\
P12 & C   & C   & C   & C   & C   \\
P13 & C   & C   & C   & C   & C   \\
P14 & C   & C   & C   & M   & C   \\
P15 & M   & M   & C   & W   & C   \\
P16 & M   & C   & W   & W   & C   \\
P17 & C   & C   & C   & C   & C   \\
P18 & C   & C   & C   & --- & C   \\
P19 & C   & C   & C   & C   & C   \\
P20 & C   & C   & C   & C   & C   \\
P21 & C   & C   & C   & C   & C   \\
P22 & --- & --- & --- & --- & --- \\
\bottomrule
\end{tabular}

\end{minipage}
\hfill
\begin{minipage}[t]{0.47\textwidth}
\vspace{0pt}
\centering

\captionof{table}{Number of problematic commits identified.}
\label{tab:score_distribution}

\footnotesize
\setlength{\tabcolsep}{5pt}
\renewcommand{\arraystretch}{0.9}

\begin{tabular}{rrr}
\toprule
Score & $n$ & $n(\%)$ \\
\midrule
0  & 6 & 27.3\% \\
1  & 0 &  0.0\% \\
2  & 1 &  4.5\% \\
3  & 4 & 18.2\% \\
4  & 2 &  9.1\% \\
5  & 1 &  4.5\% \\
6  & 0 &  0.0\% \\
7  & 1 &  4.5\% \\
8  & 0 &  0.0\% \\
9  & 2 &  9.1\% \\
10 & 3 & 13.6\% \\
11 & 1 &  4.5\% \\
12 & 1 &  4.5\% \\
13 & 0 &  0.0\% \\
\midrule
Total & 22 & 100\% \\
\bottomrule
\multicolumn{3}{l}{Mean = 4.8, Median = 3.5} \\
\end{tabular}

\vspace{1.5cm}

\captionof{table}{External information sources consulted by
participants ($n=22$). Participants may appear in more than one
category.}
\label{tab:info-sources}

\footnotesize
\setlength{\tabcolsep}{3pt}
\renewcommand{\arraystretch}{1.05}

\begin{tabular}{
    @{}
    l
    c
    >{\raggedright\arraybackslash}p{3.0cm}
    @{}
}
\toprule
\textbf{Source type}
    & \textbf{\textit{n}}
    & \textbf{Representative participants} \\
\midrule
GitHub official documentation
    & 16 & P3--P5, P7--P8, P10--P20 \\
Third-party blog/tutorial
    & 5 & P1, P2, P6, P18, P20 \\
Stack Overflow/community Q\&A
    & 3 & P7, P8, P18 \\
GitLab official documentation
    & 3 & P4, P5, P19 \\
Git-SCM documentation
    & 2 & P9, P12 \\
YouTube tutorials
    & 2 & P3, P6 \\
AI tools (ChatGPT, AI Overview)
    & 2 & P2, P17 \\
\bottomrule
\end{tabular}

\end{minipage}
\end{table*}

\begin{table*}
\centering
\caption{Descriptive statistics for the principal quantitative outcomes.
$M$ denotes arithmetic mean; $GM$ denotes geometric mean; CI denotes
confidence interval. Dashes indicate that a statistic was not applicable
or was not calculated.}
\label{tab:descriptive-results}
\small
\setlength{\tabcolsep}{7pt}
\renewcommand{\arraystretch}{1.12}
\begin{tabular}{@{}lrrrrrr@{}}
\toprule
\textbf{Outcome} &
\textbf{$n$} &
\textbf{Mean/GM} &
\textbf{SD} &
\textbf{Median} &
\textbf{IQR} &
\textbf{95\% CI} \\
\midrule
Initial-setup ASQ
    & 22 & $M=6.12$  & 0.75  & 6.00  & 1.25  & $[5.8,6.5]$ \\
Initial-setup SUS
    & 20 & $M=76.62$ & 18.30 & 80.00 & 21.25 & $[68.1,85.2]$ \\
Initial-setup time (min)
    & 20 & $GM=44.18$ & -- & 52.50 & 37.50 & -- \\
Semester-end SUS
    & 22 & $M=77.61$ & 17.45 & 81.25 & 20.62 & $[69.9,85.3]$ \\
Second-device ASQ
    & 22 & $M=5.32$ & 1.51 & 5.50 & 2.25 & $[4.7,6.0]$ \\
Second-device time (min)
    & 22 & $GM=51.04$ & -- & 42.50 & 67.50 & -- \\
Verification-task ASQ
    & 21 & $M=5.21$ & 1.45 & 5.33 & 2.33 & $[4.6,5.9]$ \\
Verification time (min)
    & 19 & $GM=39.01$ & -- & 52.50 & 43.75 & -- \\
Problematic commits identified (of 13)
    & 22 & $M=4.77$ & 4.21 & 3.50 & 8.50 & -- \\
\bottomrule
\end{tabular}
\end{table*}

\begin{table*}
\centering
\caption{Exploratory inferential analyses for initial setup, sustained
use, and second-device setup. Group summaries report $M$ and $SD$ for
SUS and ASQ and $GM$, median, and IQR for completion time. For
Mann--Whitney and Wilcoxon tests, the effect size is
$r=|Z|/\sqrt{N}$. Pearson's $r$ and Spearman's $\rho$ are themselves
effect-size estimates.}
\label{tab:inferential-adoption}
\scriptsize
\setlength{\tabcolsep}{3.5pt}
\renewcommand{\arraystretch}{1.18}
\begin{tabularx}{\textwidth}{
    @{}
    p{2.8cm}
    p{1.7cm}
    X
    l
    c
    c
    p{2.1cm}
    @{}
}
\toprule
\textbf{Analysis} &
\textbf{Sample} &
\textbf{Relevant descriptive summaries} &
\textbf{Test} &
\textbf{Statistic} &
\textbf{$p$} &
\textbf{Effect/95\% CI} \\
\midrule

Pain points and initial-setup time
&
Pain: $n=15$;
none: $n=5$
&
Pain: $GM=57.74$, median $=60.00$, IQR $=50.00$;
none: $GM=19.79$, median $=30.00$, IQR $=15.00$
&
Mann--Whitney
&
$U=66.0$
&
.013
&
$r=.554$
\\

\addlinespace

Initial-setup time and ASQ
&
$n=20$
&
See Table~\ref{tab:descriptive-results}
&
Pearson
&
$r=-.473$
&
.035
&
95\% CI $[-.757,-.038]$
\\

Initial-setup time and ASQ
&
$n=20$
&
See Table~\ref{tab:descriptive-results}
&
Spearman
&
$\rho=-.494$
&
.027
&
--
\\

\addlinespace

Initial-setup time and SUS
&
$n=19$
&
See Table~\ref{tab:descriptive-results}
&
Pearson
&
$r=-.640$
&
.003
&
95\% CI $[-.848,-.262]$
\\

Initial-setup time and SUS
&
$n=19$
&
See Table~\ref{tab:descriptive-results}
&
Spearman
&
$\rho=-.574$
&
.010
&
--
\\

\addlinespace

Initial versus semester-end SUS
&
Matched $n=20$
&
Initial: $M=76.62$, $SD=18.30$, median $=80.00$, IQR $=21.25$;
semester-end: $M=76.75$, $SD=18.10$, median $=80.00$, IQR $=23.12$
&
Wilcoxon
&
$W=83.50$
&
.930
&
$r=.020$
\\

\addlinespace

Signing consistency and semester-end SUS
&
All signed: $n=18$;
not all: $n=4$
&
All signed: $M=75.97$, $SD=18.43$, median $=81.25$, IQR $=21.88$;
not all: $M=85.00$, $SD=10.80$, median $=82.50$, IQR $=10.00$
&
Mann--Whitney
&
$U=27.50$
&
.495
&
$r=.146$
\\

\addlinespace

Initial versus second-device time
&
Matched $n=20$
&
Initial: $GM=44.18$, median $=52.50$, IQR $=37.50$;
second device: $GM=51.99$, median $=42.50$, IQR $=78.75$
&
Wilcoxon
&
$W=40.0$
&
.255
&
$r=.255$
\\

\addlinespace

Migration approach and second-device ASQ
&
Copy: $n=12$;
new: $n=10$
&
Copy: $M=5.28$, $SD=1.77$;
new: $M=5.37$, $SD=1.20$
&
Mann--Whitney
&
$U=58.0$
&
.921
&
$r=.021$
\\

\addlinespace

Migration approach and second-device time
&
Copy: $n=12$;
new: $n=10$
&
Copy: $GM=51.33$, median $=52.50$, IQR $=78.75$;
new: $GM=50.69$, median $=40.00$, IQR $=30.00$
&
Mann--Whitney
&
$U=57.0$
&
.868
&
$r=.036$
\\

\bottomrule
\end{tabularx}
\end{table*}

\begin{table*}[t]
\centering
\caption{Exploratory inferential analyses for the commit-verification
task. Detection performance is the number of problematic commits
identified out of 13. For Mann--Whitney tests, the effect size is
$r=|Z|/\sqrt{N}$. Pearson's $r$ and Spearman's $\rho$ are themselves
effect-size estimates.}
\label{tab:inferential-verification}
\scriptsize
\setlength{\tabcolsep}{3.5pt}
\renewcommand{\arraystretch}{1.18}
\begin{tabularx}{\textwidth}{
    @{}
    p{2.8cm}
    p{1.7cm}
    X
    l
    c
    c
    p{2.1cm}
    @{}
}
\toprule
\textbf{Analysis} &
\textbf{Sample} &
\textbf{Relevant descriptive summaries} &
\textbf{Test} &
\textbf{Statistic} &
\textbf{$p$} &
\textbf{Effect/95\% CI} \\
\midrule

Verification time and detection
&
$n=19$
&
See Table~\ref{tab:descriptive-results}
&
Pearson
&
$r=.668$
&
.002
&
95\% CI $[.306,.861]$
\\

Verification time and detection
&
$n=19$
&
See Table~\ref{tab:descriptive-results}
&
Spearman
&
$\rho=.631$
&
.004
&
--
\\

\addlinespace

Verification ASQ and detection
&
$n=21$
&
See Table~\ref{tab:descriptive-results}
&
Pearson
&
$r=-.081$
&
.726
&
95\% CI $[-.496,.363]$
\\

Verification ASQ and detection
&
$n=21$
&
See Table~\ref{tab:descriptive-results}
&
Spearman
&
$\rho=-.137$
&
.553
&
--
\\

\addlinespace

Earlier signing mechanism and detection
&
GPG: $n=16$;
SSH: $n=6$
&
GPG: $M=5.06$, $SD=4.36$, median $=3.50$, IQR $=7.00$;
SSH: $M=4.00$, $SD=4.05$, median $=3.50$, IQR $=6.00$
&
Mann--Whitney
&
$U=55.0$
&
.627
&
$r=.104$
\\

\addlinespace

Signer-file configuration and detection
&
Configured: $n=8$;
not configured: $n=14$
&
Configured: $M=9.12$, $SD=2.53$, median $=10.00$, IQR $=1.75$;
not configured: $M=2.29$, $SD=2.61$, median $=2.50$, IQR $=3.00$
&
Mann--Whitney
&
$U=108.0$
&
$<.001$
&
$r=.760$
\\

\addlinespace

Giving up and detection
&
Gave up: $n=6$;
did not: $n=16$
&
Gave up: $M=1.50$, $SD=1.64$, median $=1.50$, IQR $=3.00$;
did not: $M=6.00$, $SD=4.24$, median $=6.00$, IQR $=7.25$
&
Mann--Whitney
&
$U=18.0$
&
.027
&
$r=.470$
\\

\bottomrule
\end{tabularx}
\end{table*}

\end{document}